\documentclass[11pt, twocolumn, superscriptaddress, aps, prl]{revtex4-2}
\usepackage{amsmath, amssymb, graphicx, hyperref, xcolor}
\usepackage{csquotes}
\usepackage{comment}
\usepackage{braket}
\usepackage{natbib}
\usepackage{url}
\usepackage{titlesec}
\usepackage{geometry}
\usepackage{csquotes}
\usepackage{caption}
\usepackage{hyperref}
\usepackage{setspace} 
\usepackage{subcaption}
\singlespace

\titleformat{\section}
  {\normalfont\fontsize{12}{15}\bfseries}{\thesection}{1em}{}
\titleformat{\subsection}
  {\normalfont\fontsize{10}{15}\bfseries}{\thesubsection}{1em}{}
\begin{document}

\title{Quantum Error Mitigation via Linear-Depth Verifier Circuits}
\author{A. Mingare}
\affiliation{Centre for Computational Science, Department of Chemistry, University College London, Gower Street, London, WC1E 6BT, United Kingdom}
\author{A. Moroz}
\affiliation{Department of Computer Science, University College London, Gower Street, London, WC1E 6BT, United Kingdom}
\author{M. D.  Kovács}
\affiliation{Department of Physics and Astronomy, University College London, Gower Street, London, WC1E 6BT, United Kingdom}
\author{A. G. Green }
\affiliation{Department of Physics and Astronomy, University College London, Gower Street, London, WC1E 6BT, United Kingdom}

\date{\today}

\maketitle       
\textit{
Implementing many important sub-circuits on near-term quantum devices remains a challenge due to the high levels of noise and the prohibitive depth on standard nearest-neighbour topologies. Overcoming these barriers will likely require quantum error mitigation (QEM) strategies. This work introduces the notion of efficient, high-fidelity verifier circuit architectures that we propose for use in such a QEM scheme. We provide a method for constructing verifier circuits for any quantum circuit that is accurately represented by a low-dimensional matrix product operator (MPO). We demonstrate our method by constructing explicit verifier circuits for multi-controlled single unitary gates as well as the quantum Fourier transform (QFT). By transpiling the circuits to a 2D array of qubits, we estimate the crossover point where the verifier circuit is shallower than the circuit itself, and hence useful for QEM. We propose a method of \textit{in situ} QEM using the verifier circuit architecture. We conclude that our approach may be useful for calibrating quantum sub-circuits to counter coherent noise but cannot correct for the incoherent noise present in current devices.
}

\section{Introduction}

The capability of the quantum Fourier transform (QFT) to detect periodicities in quantum states has rendered it an important sub-circuit in many quantum algorithms, such as Shor’s algorithm for integer factorisation \cite{Shor1997} and quantum phase estimation \cite{Kitaev1995}. Similarly, multi-controlled unitary (MCU) gates are essential components in several promising quantum algorithms including Grover's algorithm \cite{grover1996fast} and quantum dynamics simulations, particularly the linear combination of unitaries method \cite{wiebe2012hamiltonian}. Despite the importance of these sub-circuits for the algorithms mentioned, current noisy intermediate-scale quantum (NISQ) computing hardware prohibits their immediate use. When executed on NISQ hardware these circuits have very low fidelities. This is due to the need for fine rotation control and the quadratic-scaling depth when transpiled to nearest-neighbour topologies. Thus, it is essential to adapt such quantum sub-circuits to NISQ hardware. 

Quantum error mitigation (QEM) strategies aim to extract useful results from NISQ hardware by accounting for noise. Most QEM strategies depend on post-processing information to correct for errors on the level of the entire circuit \cite{cai2023quantum}. However, it would be useful to design QEM strategies to optimise the performance of specific, important sub-circuits that can then be embedded into larger algorithms. 

The QEM scheme proposed in this work was inspired by several key results regarding the QFT. Initially, the singular values of the QFT's Schmidt decomposition were found to be uniform, implying maximum operator entanglement \cite{Nielson2003, Tyson2002}. However, later numerical evidence indicated that this maximum entanglement was entirely due to the classically simulable bit-reversal section of the QFT \cite{Woolfe2017}. By providing analytical bounds on the entanglement of the QFT, \textit{Chen et al.} were able to construct an efficient and accurate matrix product operator (MPO) representation of this section of the QFT, with a bond dimension of $8$ for all qubit numbers \cite{Chen2022}. We find that the structure of this QFT MPO naturally lends itself to that of a \textit{verifier circuit}, which we define in Section II. Our results indicate that the verifier circuit construction scales linearly with the number of qubits, even on nearest-neighbour topologies. Furthermore, we realise any quantum sub-circuit that can be accurately represented by a low bond dimension MPO admits a linear-depth verifier circuit. In addition to the QFT, this class of circuits also includes MCUs. We demonstrate the utility of the verifier circuit construction in developing in situ QEM schemes to improve the fidelity of important quantum sub-circuits. 

\section{Verifier Circuit Construction}

Tensor networks are a useful tool for simulating quantum systems \cite{Biamonte2017}. Any quantum circuit can be represented by an MPO using the \enquote*{zip-up} algorithm described in \cite{Chen2022}. This method iteratively performs tensor contractions and singular value decompositions (SVDs) on the quantum gates constituting the quantum circuit to form the MPO. To find a low-dimensional representation, the bond-dimension of tensors is truncated to a fixed small value retaining only the largest singular values in the SVD sweeps. For the class of classically simulable quantum circuits, the bond truncated MPO is a high-fidelity representation of the original quantum circuit. Many important quantum sub-routines fall into this class. The results of Woolfe et al. \cite{Woolfe2017} and Chen et al. \cite{Chen2022} show that a bond dimension of $8$ is sufficient to faithfully represent the QFT circuit ($>99$\% fidelity). MCU gates with a single target qubit are perfectly represented by an MPO with bond dimension of just $2$. 

The output of the zip-up algorithm produces a MPO that is unitary from top to bottom. This indicates that the MPO has a natural interpretation as a \textit{verifier circuit}. A verifier circuit for a quantum circuit $U$ takes as input an arbitrary product state $\ket{\psi} \otimes \ket{\psi'}$ and outputs a known result with very high fidelity if and only if $\ket{\psi'} = U \ket{\psi}$. Therefore a verifier circuit is able to check whether an implementation of a quantum circuit has high fidelity. To convert an MPO to a verifier circuit, the dimension of the input legs must be made equal to the dimension of the output legs so that the tensors can be interpreted as quantum gates. To achieve this, the bond dimension is not truncated on the final SVD sweep during the zip-up algorithm. Rather, at each tensor, we retain the full unitary obtained from the SVD but only treat the truncated subspace corresponding to the largest singular values as an internal bond of the tensor network. The remaining subspace is associated to an external leg so that bond truncation is replaced by a post selection of $\ket{0}$ on that external leg. Mathematically this can be thought of as projecting those qubits onto the $\ket{0}$ subspace. The final tensor in the MPO is replaced by the unitary part of its SVD while the non-unitary part becomes the expected output of the verifier circuit after post-selecting all the external bonds on $\ket{0}$. The verifier circuit construction is depicted in Figure 1.

\begin{figure*}[ht]
    \centering
    \includegraphics[width=\linewidth]{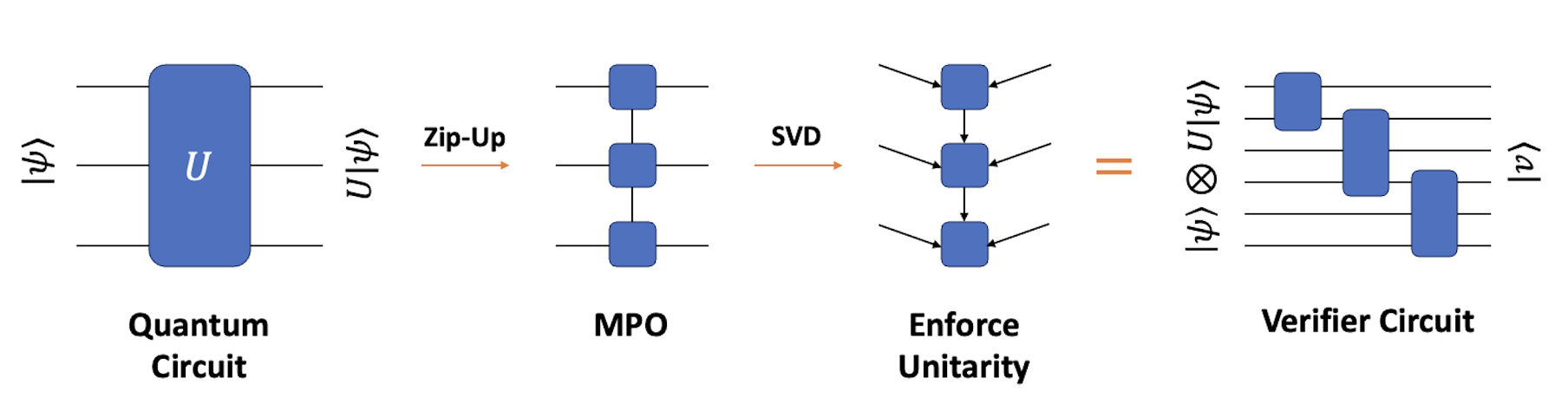}
    \caption{Verifier circuit construction for a quantum circuit $U$. The zip-up algorithm is used to convert the circuit into a MPO. Square unitarity is enforced with a final SVD sweep. The resulting MPO has a natural interpretation as a verifier circuit. This circuit takes as input an arbitrary state $\ket{\psi}$ and its transform $U\ket{\psi}$. Upon post-selecting on all $\ket{0}$ states, the output state $\ket{v}$ of the final gate is known.}
    \label{fig:verifier_circuit_construction}
\end{figure*}

\section{Verifier Circuit Scaling}

By construction, the depth of a verifier circuit scales linearly with the number of qubits and contains only local entangling gates. Therefore when transpiled to real hardware on a 2D lattice, the depth remains linear. This is an improvement over the quadratic scaling required for a naive verifier circuit on real hardware (namely, the inverse circuit). Furthermore, the verifier circuit being shallower than the original circuit is essential for in situ QEM as will be discussed in the next section. 

By transpiling to a 2D array of qubits, we can estimate the crossover point where the verifier circuit becomes shallower than the original circuit. This is shown in Figure 2 for the QFT circuit and in Figure 3 for the MCX circuit.

\begin{figure}
\includegraphics[width=\linewidth]{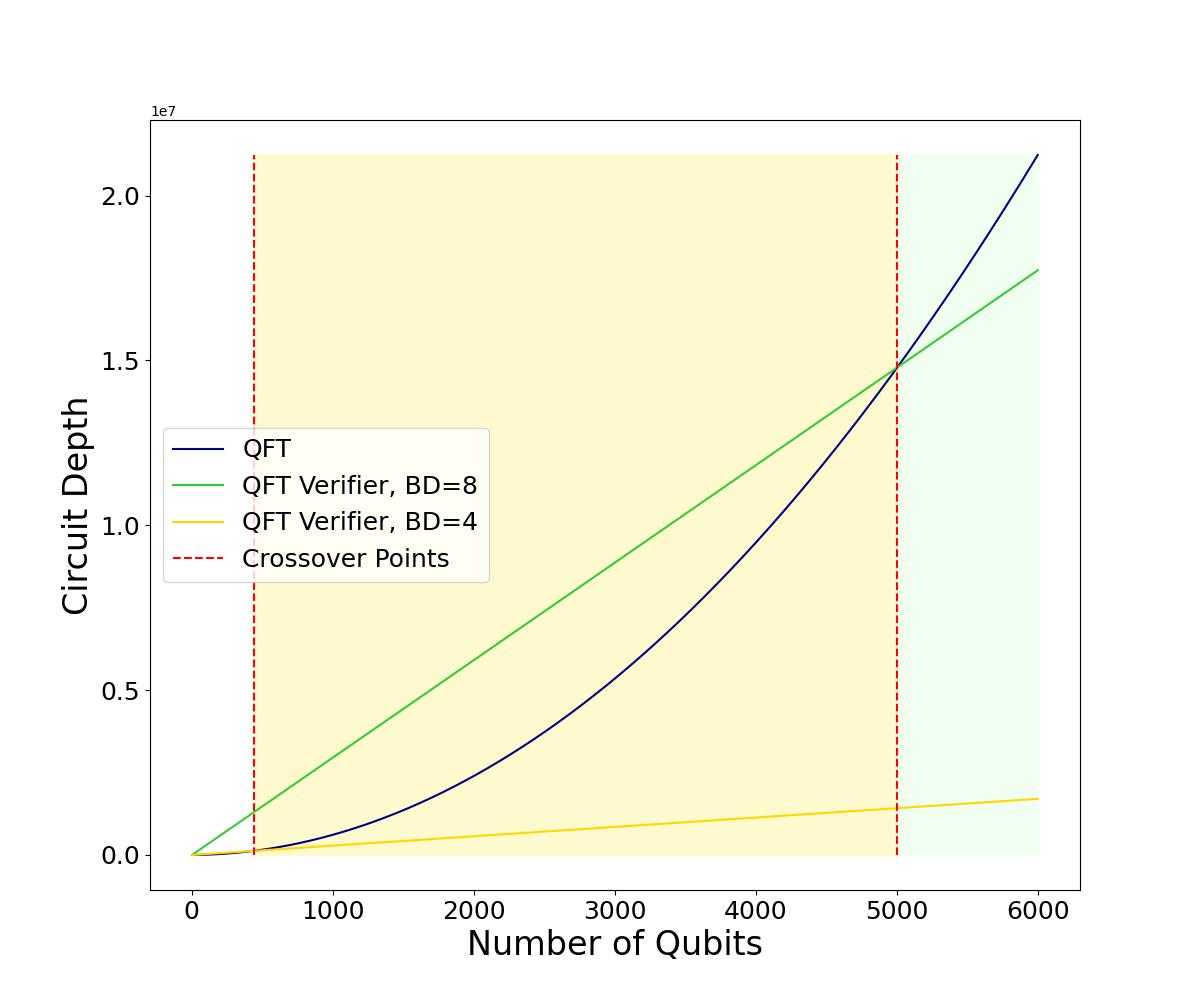}
\caption{The plot shows the depth of the QFT circuit (blue), the depth of the QFT verifier circuit with bond dimension (BD) $8$ (green), and BD $4$ (yellow). The crossover points at $5003$ qubits and $440$ qubits are indicated by vertical dashed red lines.}
\label{fig:qft_crossover}
\end{figure}
\begin{figure}
\includegraphics[width=\linewidth]{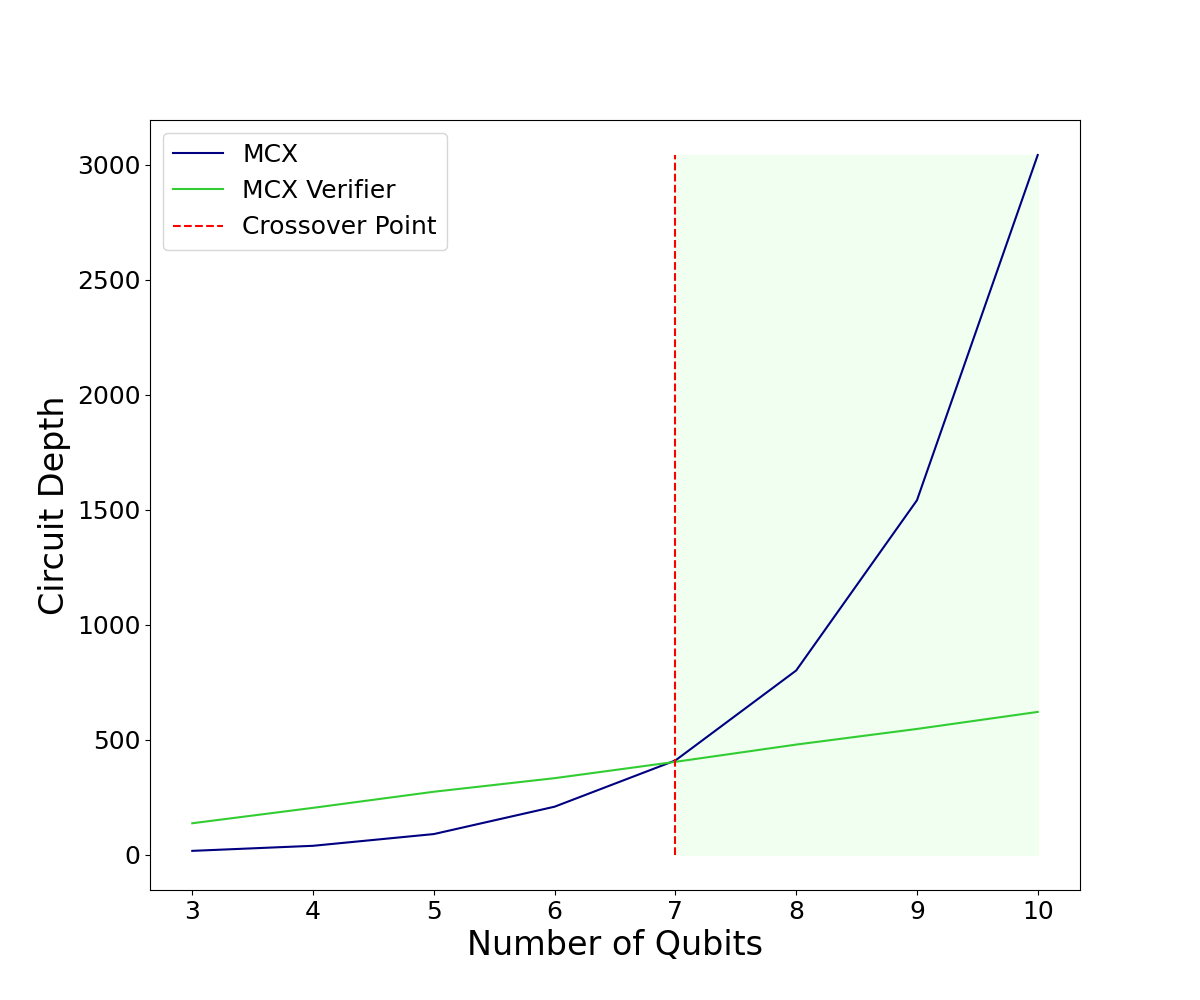}
\caption{The plot shows the depth of the MCX circuit (blue) and the depth of the MCX verifier circuit (green). The crossover point at $7$ qubits is indicated by a vertical dashed red line.}
\label{fig:mcx_crossover}
\end{figure}

We estimate that the crossover point where the QFT verifier circuit is shallower than the QFT occurs at $5003$ qubits, well beyond any current or near-term NISQ devices. The crossover point is so high because for a maximum bond dimension of $8$ in the QFT MPO, the resulting verifier circuit is built out of arbitrary $5$-qubit unitary gates which are very deep when transpiled to a native gate set involving single-qubit unitaries and a two-qubit entangling gate. One way to reduce this transpilation overhead is to reduce the maximum bond dimension of the QFT MPO to $4$ so that the verifier circuit is built out of $4$-qubit unitary gates. This sacrifices a small amount of fidelity for a saving in the overall circuit depth. Doing so, the crossover point is reduced by an order of magnitude from $5003$ to $440$ as shown in Figure \ref{fig:qft_crossover}. Further reducing the maximum bond dimension of the QFT MPO to $2$ brings the crossover point down by another order of magnitude to $52$ however the trade-off in fidelity becomes potentially prohibitively large. 

However, because the MCX circuit is perfectly represented by a MPO with bond dimension $2$, the resulting verifier circuit is much simpler and the crossover point is as low as $7$ qubits, well within the realm of near-term devices.

\section{Quantum Error Mitigation with Verifier Circuits}

The verifier circuit construction can verify the action of a quantum circuit on an arbitrary input state which is essential for calibration. We therefore propose that our linear depth verifier circuit architecture may be useful in developing in situ error mitigation schemes for quantum circuits on real hardware. We propose two approaches for this.

The first approach shown in Figure 4 involves first drawing a state randomly from a uniform distribution of trial states and preparing it twice. One of the prepared states is acted on by the circuit $U$ before going through an error mitigating layer of parameterised unitary gates, $E(\theta_i)$. This state together with its copy are then passed to the verifier circuit, $U_{\mathrm{verifier}}$. The error mitigating gates are iteratively optimised to maximise the fidelity of the verifier circuit with respect to the expected output that we learn during the construction of the verifier circuit.

The second approach shown in Figure 5 is similar, however instead of an error mitigating layer of unitaries, the rotation angles of the gates involved in the circuit $U$ are themselves treated as variational parameters that are optimised using the output fidelity of the verifier circuit. 
\begin{figure}[ht]
\centering
\includegraphics[width=\linewidth]{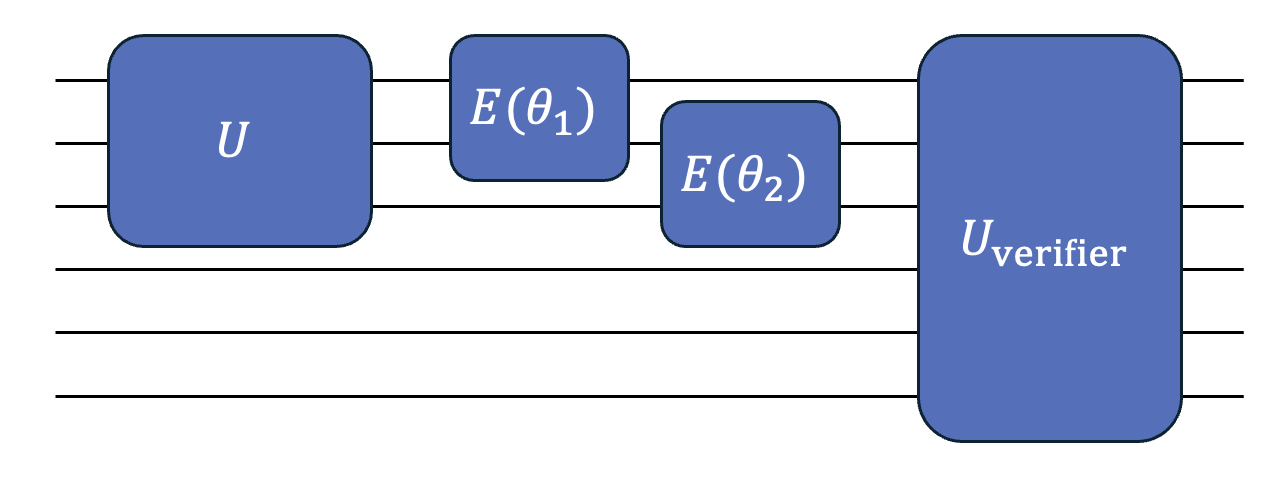}
\caption{A quantum error mitigation scheme using the verifier circuit architecture. The error mitigating gates $E(\theta_i)$ are optimised to maximise the fidelity of the output of the verifier circuit.}
\end{figure}

\begin{figure}[ht]
\centering
\includegraphics[width=0.6\linewidth]{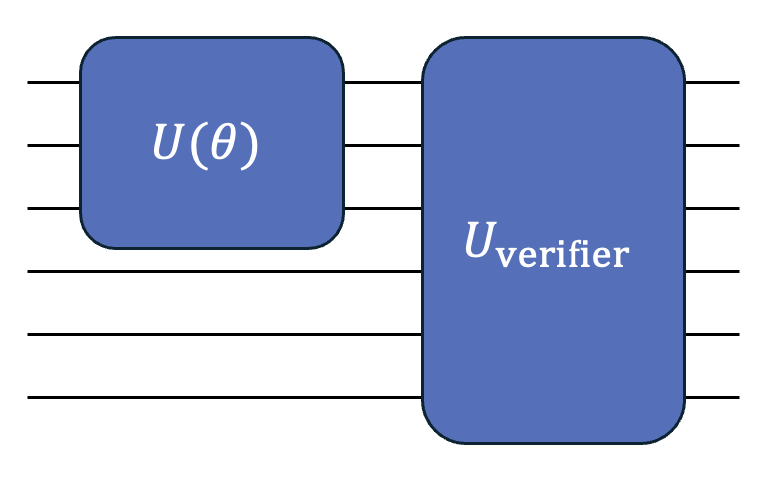}
\caption{A quantum error mitigation scheme using the verifier circuit architecture. The rotation angles comprising the circuit $U$ are treated as variational parameters and optimised to maximise the fidelity of the output of the verifier circuit.}
\end{figure}
These error mitigation schemes are similar in flavour to the accreditation protocol of \cite{ferracin2019accrediting}. The accreditation protocol uses circuits of equal size to the target circuit to estimate the probability that an error occurs. This was extended in \cite{mezher2022mitigating} to a protocol to find error-mitigated expectation values of quantum circuits. In contrast, the strategy presented here is via a linear depth verifier circuit that achieves error mitigation at an operator level. Hence it is more general and allows the error-mitigated circuit to be embedded in larger routines. 

These strategies will be feasible when the depth of the verifier circuit is significantly less than the depth of the circuit itself so that we can assume the majority of errors are coming from the original circuit and can be successfully mitigated. As shown in the previous section, the verifier circuit for the MCX gate becomes more shallow than the MCX itself at just $7$ qubits indicating that a QEM sheme based on the verifier circuit construction may be possible on near-term devices. The demonstrations in the following sections will use the MCX gate and its verifier circuit. 

\subsection{Coherent Errors}

An in situ QEM strategy should at a minimum correct for coherent control errors (that is, device calibration). These arise from systematic and random over/under rotations of the target qubits. Using a simplified noise model that only includes these sources of noise we are able to optimise the performance of a noisy MCX gate. For demonstration purposes we consider a $C^2X$ gate and treat the verifier circuit as noiseless to mimic the scenario where the verifier circuit is significantly more shallow than the original circuit. 

After decomposing $C^2X$ as a series of one- and two-qubit rotations, the rotation angles are treated as noisy parameters where the noise is sampled from a Gaussian distribution with non-zero mean. No other noise sources are included in this simulation. Table 1 shows the fidelity of the noisy $C^2X$ circuit as well as the fidelity following calibration using both of our proposed QEM schemes. Each fidelity is calculated using
\begin{equation}
    \mathcal{F} = \frac{\mathrm{Tr}(U_{\mathrm{noisy}}^\dagger U_{\mathrm{ideal}})}{\mathrm{Tr}(U_{\mathrm{ideal}}^\dagger U_{\mathrm{ideal}})},
\end{equation}
\begin{table}[h!]
\centering
 \begin{tabular}{||c c||} 
 \hline
 QEM Scheme & $C^2X$ Fidelity  \\ [0.5ex] 
 \hline\hline
 None & 0.788  \\ 
 Method 1 & 0.872  \\
 Method 2 & 0.990 \\
 \hline
 \end{tabular}
\caption{The fidelity of $C^2X$ following each QEM scheme.}
\end{table}
Both methods significantly increase the fidelity of $C^2X$ demonstrating that the verifier circuit architecture provides a useful metric for device calibration on the sub-circuit level. 

Method 2 performs better than Method 1. However, the number of parameters to optimise using Method 1 will scale linearly with the size of the circuit whereas Method 2 will scale quadratically for both MCX and QFT circuits. Therefore, Method 1 may be more appropriate for calibrating larger sub-circuits.

\subsection{Incoherent Errors}

Our proposed QEM scheme can only provide unitary corrections. We demonstrate here that following device calibration the dominant sources of noise are incoherent and unable to be corrected with unitary gates. 

Using a realistic noise model, including non-unitary errors such as thermal relaxation, based on IBM's Hanoi quantum processor, the initial fidelity of a $C^5X$ gate is just $0.754$. An optimal correcting process $E$ to this gate should satisfy
\begin{equation}
    E {C^5X}_{\mathrm{noisy}} = C^5X,
\end{equation}
where ${C^5X}_{\mathrm{noisy}}$ denotes the average noisy implementation of the gate. By inverting this process we can isolate $E$. The optimal unitary correction to ${C^5X}_{\mathrm{noisy}}$ is then the closest unitary to $E$, denoted $U_E$ which can be found by the polar decomposition. Unfortunately, the fidelity of $U_E {C^4X}_{\mathrm{noisy}}$ compared to $C^5X$ is just $0.759$. From this we conclude that any remaining coherent errors after the device has been calibrated are almost negligible when compared to incoherent errors present in near-term devices. 

\section{Conclusion}

In this paper, we have studied efficient verifier circuit architectures for multi-controlled unitaries and the quantum Fourier transform. We presented a general method for constructing such a circuit for any quantum algorithm that has an accurate representation by a low-dimensional matrix-product operator. We investigated the utility of our construction within quantum error mitigation schemes and found a use-case for device calibration in the presence of coherent error. 

Possible avenues for further work include investigating which other quantum subroutines admit a low bond-dimension MPO representation and overcoming the limitation of our approach in the presence of incoherent noise. The latter would require developing different approaches to in situ QEM such as by using stochastic error mitigating gates rather than static unitaries. 

\pagestyle{plain}
\bibliography{bib}

\end{document}